\documentclass[reprint,amsmath,amssymb,aps,prapplied]{revtex4-1}

\usepackage{graphicx}
\usepackage{dcolumn}
\usepackage{bm}
\usepackage{gensymb}
\usepackage{graphicx}
\usepackage{multirow}
\usepackage[usenames]{color}
\usepackage{dcolumn}
\usepackage{bm}
\usepackage{hyperref}
\usepackage[mathlines]{lineno}

\hypersetup{
	colorlinks = true,
	urlcolor   = blue,
	linkcolor  = blue,
	citecolor  = blue
}

\begin{document}

\preprint{APS/123-QED}

\title{Reconfigurable photonics on a glass chip}

\author{I.V.Dyakonov}
\email{iv.dyakonov@physics.msu.ru}
\author{I.A.Pogorelov, I.B.Bobrov, A.A.Kalinkin, S.S.Straupe, S.P.Kulik}
\affiliation{Faculty of Physics, M. V. Lomonosov Moscow State University and \\ 
	Quantum Technologies Centre, M. V. Lomonosov Moscow State University, Leninskie Gory 1, Moscow, Russia, 119991}
\author{P.V.Dyakonov}
\affiliation{Skobeltsyn Institute of Nuclear Physics, M. V. Lomonosov Moscow State University, Leninskie Gory 1, Moscow, Russia, 119991}
\author{S.A.Evlashin}
\affiliation{Center for Design, Manufacturing and Materials, Skolkovo Institute of Science and Technology, 3 Nobel Street, Moscow, 143026, Russia}

\date{\today}

\begin{abstract}
Reconfigurability of integrated photonic chips plays a key role in current experiments in the area of linear-optical quantum computing. We demonstrate a reconfigurable multiport interferometer implemented as a femtosecond laser-written integrated photonic device. The device includes a femtosecond laser-written $4\times 4$ multiport interferometer equipped with 12 thermooptical phase shifters, making it a universal programmable linear-optical circuit. We achieve a record fast switching time for a single nested Mach-Zender interferometer of $\sim10$~ms and quantitatively analyse the reconfigurability of the optical circuit. We believe, that our results will improve the current state of quantum optical experiments utilizing femtosecond laser-written photonic circuits.
\end{abstract}

\pacs{Valid PACS appear here}
\maketitle


\section{Introduction}

The quest for building a large-scale linear optical quantum computer implies the development of miniature, stable and precise optical components. A remarkable success of the microelectronic industry proves the immense efficiency of integrated device fabrication and dictates the future trend for quantum photonic device engineering. Since the pioneering work \cite{OBrien2008} an integrated photonic approach firmly keeps the leadership in precision quantum optical experiments \cite{Silberberg2008, Shadbolt2011}, optical quantum computing \cite{OBrien2015} and quantum simulation \cite{Harris2017}. Recent works \cite{Silverstone2016} have also demonstrated the potential of adopting standard silicon photonic technologies for quantum applications, ensuring compatibility with modern CMOS fabrication lines.

Along with stability and precision, the integrated photonic technology provides a toolset for reconfigurable circuit fabrication, endowing the experimenter to perform numerous experiments with a single device. Recently quantum computing experiments have been demonstrated on reconfigurable platforms\cite{Peruzzo2014, Paesani2017, Wang2017}, proving their potential for realizing completely different experimental settings with a single device \cite{OBrien2015}. Reconfigurability provides enough freedom to augment linear optical quantum computing experiments with machine learning algorithms \cite{Paesani2017, Wang2017}, paving the way to more sophisticated applications for the devices of increased complexity.

Currently the most advanced reconfigurable integrated photonic chips are fabricated using lithographic technology \cite{Silverstone2016} providing the best possible precision of elementary components \cite{Wilkes16}, miniaturization \cite{Piggott2015} and the fastest low-loss switching \cite{Harris2014, Lu2016}. Several different optical architectures have been realized using this technology \cite{OBrien2015, Harris2017, Ribeiro2016, Prez2017, Tanizawa2015}. Furthermore, tunable optical integrated devices are designed and fabricated for many purposes such as optical phased arrays \cite{Sun2013}, optical interconnects for microprocessors \cite{Sun2015}, signal processing \cite{Liu2016} and others. 

Recently the femtosecond laser writing technology (FSLW) has established itself as a flexible tool for rapid prototyping of integrated photonic circuits, which is especially valuable for laboratory experiments, where the fabrication time is crucial. FSLW provides low-loss waveguide writing regimes for a wide variety of wavelengths from the visible to the telecom range \cite{Szameit2009, Flamini2015, Atzeni2018}. Waveguide fabrication is possible in glasses, crystals, nonlinear materials, etc \cite{DellaValle2009}. FSLW enables polarization state manipulation capabilities \cite{Corrielli2014, Dyakonov2017} and essentially 3D waveguide circuit design \cite{Flamini2017}. The FSLW technology in the current quantum optical scope provides a versatile tool to process complex quantum states of light encoded in different degrees of freedom on the integrated photonic platform. For example, it has been applied for characterisation of hyperentangled path-polarization states \cite{Ciampini2016}, and further development of the active FSLW technology may significantly contribute to the on-chip manipulation of such states. One of the key advantages of this technology in comparison with lithography is its flexibility and very fast and inexpensive technological process.

\begin{figure*}[htbp]
    \centering
    \includegraphics[width=\linewidth]{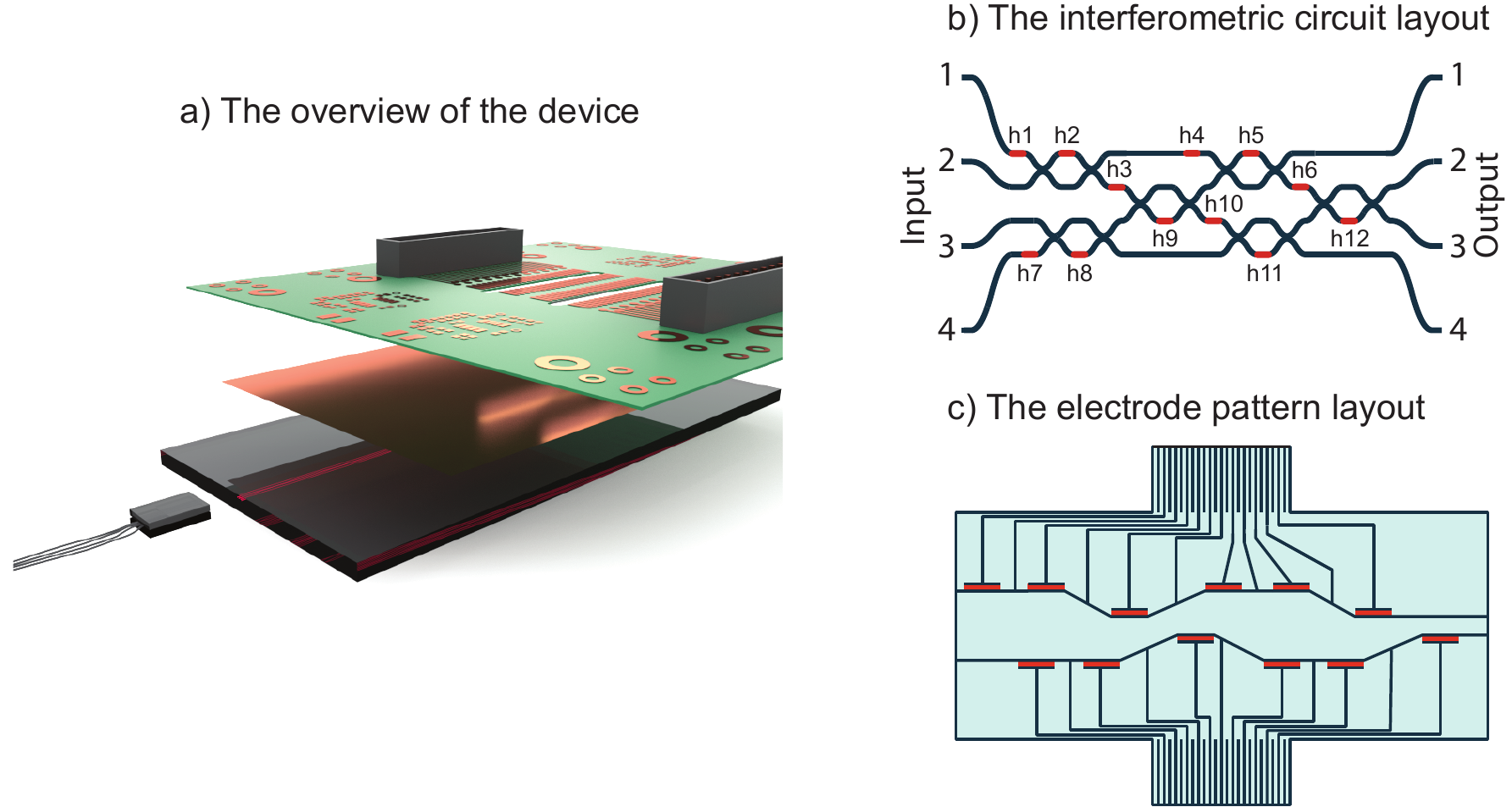}
    \caption{(a) An overview of the fabricated device. (b) A schematic of the waveguide circuit written in the chip. For geometry details please refer to the Appendix. (c) An illustration of an electric pattern engraved in the metal film sputtered on the top surface of the chip. The drawing is not to scale.}
    \label{fig:chip_model}
\end{figure*}

Active thermooptically adjustable elements have recently been introduced to the FSLW fabrication process \cite{Flamini2015, Chaboyer2017}, and several quantum optical experiments have been performed with reconfigurable laser-written circuits \cite{Ciampini2016, Crespi2017}. However, a fully reconfigurable device, capable of realizing universal unitary transformations has not been demonstrated so far. To the best of our knowledge, all previously demonstrated devices included at most two consecutive thermooptical phase-shifters \cite{Crespi2017}. A modular architecture was suggested recently to enable assembly of such larger universal interferometers form smaller modules \cite{Mennea2018}, but the electrode imprinting process used still required lithographic techniques. In this work we demonstrate the fabrication process of a fully reconfigurable multiport interferometer using FSLW only. The electrodes are engraved using the same facility as the waveguides and the only step performed outside the optical laboratory environment is the metallization of the chip surface. This approach was pioneered in \cite{Flamini2015}, but we significantly improve it to fit more thermo-optical elements on a single sample. Our device is capable of realizing a universal SU(4) unitary transformation, which is the largest unitary realized with a fully-FSLW chip so far. Moreover, we improve the switching time of the phase-shifters by two orders of magnitude reaching the level of 10~ms. The performance of the device including cross-talks between heaters is thoroughly characterized and an adaptive algorithm is introduced to tune the unitary for the desired output.

\section{Fabrication\label{fabrication}}

\begin{figure*}[htbp]
    \centering
    \begin{minipage}{0.33\textwidth}
        \centering
        \includegraphics[width=\linewidth]{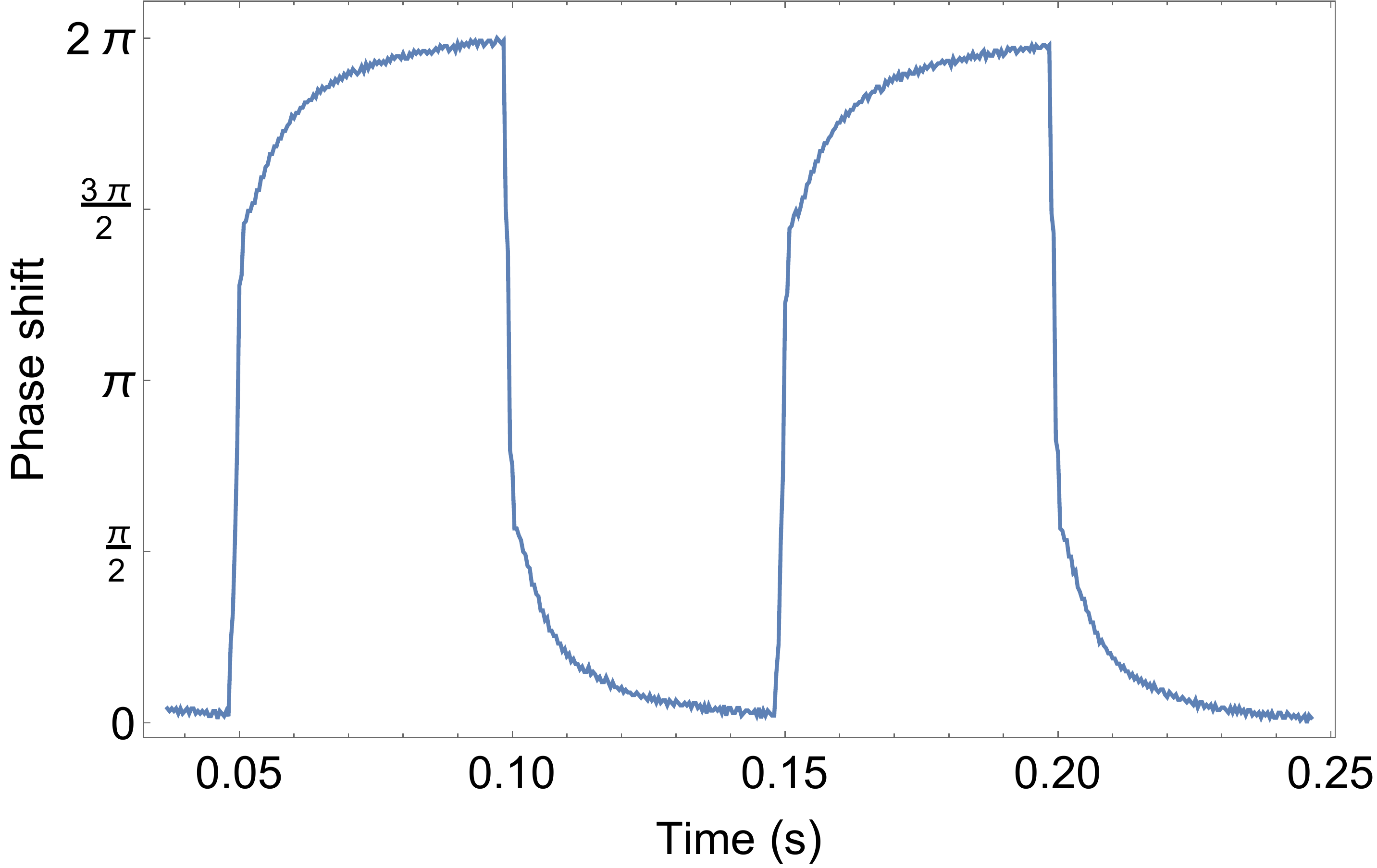}
        (a)
    \end{minipage}\hfill
    \begin{minipage}{0.33\textwidth}
        \centering
        \includegraphics[width=\linewidth]{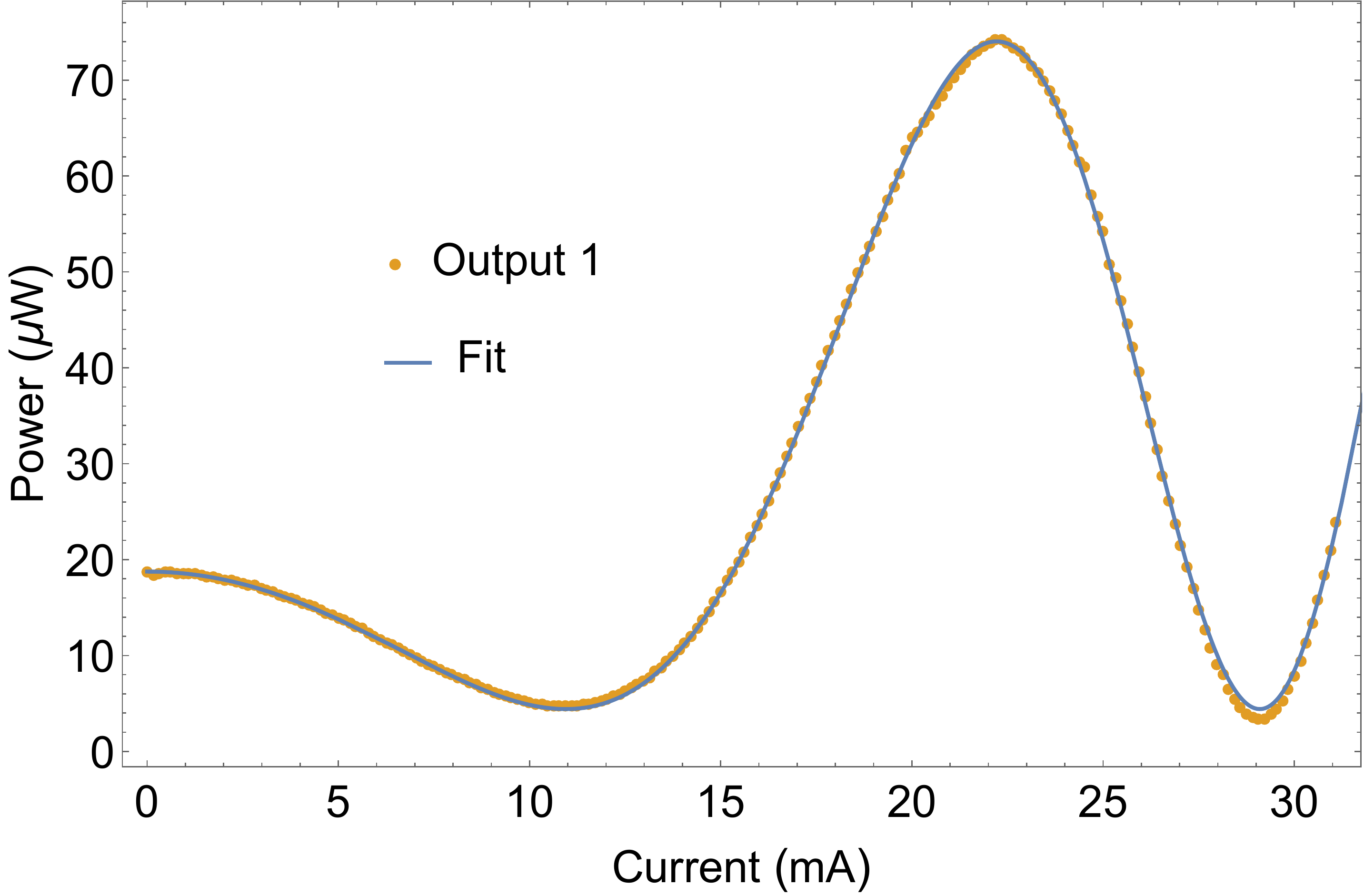}
        (b)
    \end{minipage}\hfill
    \begin{minipage}{0.33\textwidth}
        \centering
        \includegraphics[width=\linewidth]{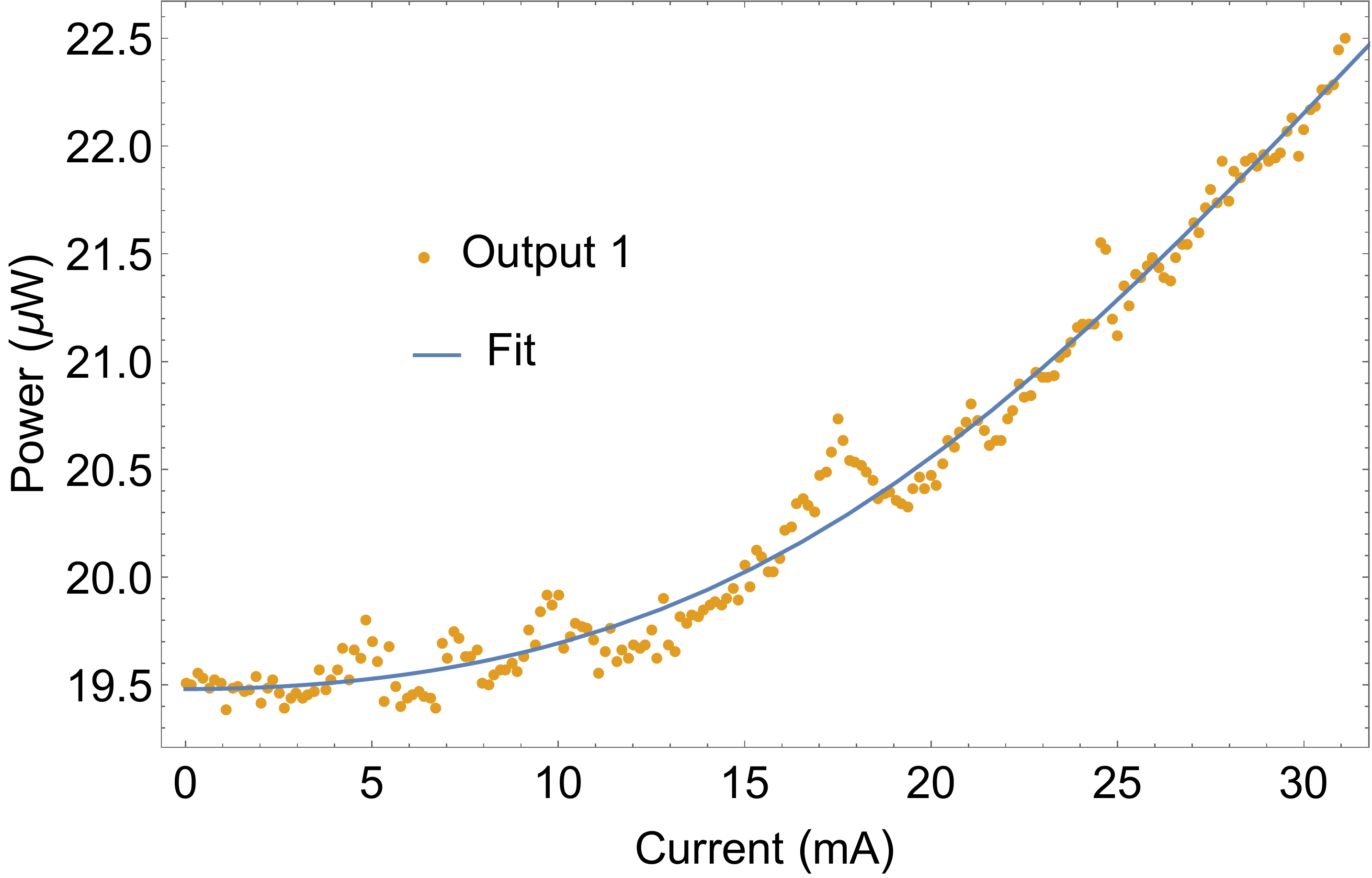}
        (c)
    \end{minipage}
    \caption{(a) The response curve for the rectangular excitation current pulse applied to the heater h8. (b) The calibration curve for the h8 heater. The interference pattern on the output port 3 was measured while sweeping the control current through the heater h8. (c) The observed cross-talk effect on the output port 1 while sweeping the control current through the heater h1.}
    \label{fig:time}
\end{figure*}

We designed the optical circuit of the reconfigurable chip using a unitary matrix factorization algorithm developed in \cite{Clements2016}. The key benefits of this elaborated procedure compared to the well-known Reck design \cite{Reck1994} are the more compact footprint and the symmetric effect of propagation loss throughout the beamsplitter mesh. The latter feature endows the symmetric beamsplitter layout with greater robustness to loss which allows fabricating optical circuits of larger depth without a drastic fidelity reduction. To achieve reconfigurability in the proposed design, one needs to tune the transmission ratio of every beamsplitter and the relative phaseshifts between the different arms of the circuit (see Fig.~\ref{fig:chip_model}). To satisfy these requirements the beamsplitters are replaced with Mach-Zender interferometers with output intensity and phase distribution adjustment enabled by thermooptical elements. The full topology of the optical circuit is shown in Fig.~\ref{fig:chip_model}(b).

The integrated photonic circuit is fabricated with a femtosecond laser writing technology. We expose a fused silica sample (100x50x5~mm, JGS1, AGOptics) to 178 nJ laser pulses (400~fs duration) at 3 MHz repetition rate. We use a second harmonic of an ytterbium fiber femtosecond laser system (Menlo Systems Bluecut). The laser light at 515~nm central wavelength is focused with a 0.55~NA aspheric lens 20~$\mu$m below the surface. Smooth waveguiding structures are inscribed inside the volume of the sample by translating it along the desired trajectory at the constant feed rate of 0.01~mm/s with an air-bearing translation stage (AeroTech FiberGlide 3D). The details of the experimental setup can be found in \cite{Dyakonov2017}. The fabricated waveguides exhibit 0.8 dB/cm propagation loss. The inscribed structures exhibit a rather low refractive index contrast, thus we set the bending radius of the curved interferometer sections to be 80 mm to ensure negligible additional bending loss. This drawback is typical for all FSLW-fabricated waveguides and severely limits the scalability of the integrated optical circuits created with this technology. Therefore a 4-mode universal interferometer circuit is a maximal universal multiport we could possibly fit to our samples. The geometrical parameters of the interferometer are summarized in Table~\ref{tab::geometry} of the Appendix. The integrated circuit is written 20~$\mu$m below the surface to ensure a fast thermooptic response. The quality of the optical waveguides written at small depth is significantly affected by the surface thickness variance. To eliminate that factor we used 5~mm thick slabs with the surface polished up to $\frac{\lambda}{10}$~@~633 nm flatness quality. After the optical circuit is written, the location of the interferometer is marked with alignment patterns. We use these markers to realign the position of the sample with the translation stage coordinate frame to engrave the heating elements at the required locations relatively to the interferometer. The position of the markers can be restored with 1~$\mu$m precision. Lastly the end faces of the chip are optically polished.

To create the heating elements on the chip we deposit a metal NiCr film via a magnetron sputtering process. The sample is washed in an ultrasonic bath in an oil removal solution and cleaned in ethanol afterwards. It is preheated up to 150~\degree C to enhance the adhesive properties of the surface and a 1~$\mu$m thick NiCr film is sputtered on the top surface of the sample. Next, we mount the chip back in the FSLW facility, realign it with the positioning stage coordinate frame using the marker pattern and engrave the electrode circuit with 50~nJ pulses at 1~MHz repetition rate focused with a 0.7~NA objective. The active heating parts of the fabricated electrodes are 3~ mm long and 50~$\mu$m wide. The full electrode pattern engraved is shown in Fig.~\ref{fig:chip_model}(c). An electrical interface with a homebuilt 12-bit digital constant current source is achieved via a PCB contacting the electrodes on the optical chip via thin metal springs. The whole assembly is mounted on the aluminum heat sink and the temperature of the chip is stabilized at 18~\degree C. The overview of the device assembly is given in Fig.~\ref{fig:chip_model}(a).

\section{Device characterization}

We tested the device using a simple setup: laser light at 808~nm was injected in the chip through a butt-coupled single-mode v-groove fiber array, light on the output was collected with a multimode v-groove fiber array and sent to fiber-coupled photodiodes. The device performance is characterized by precision of preparing the desired intensity distribution on the output, and by temporal switching constants: $t_{1}$ -- the switching time of a single interferometer and $t_{2}$ -- the time delay required for the whole system of electrodes to reach the equilibrium state. To measure the $t_{1}$ time we applied a rectangular current pulse to the heater h8 (see Fig.~\ref{fig:chip_model}(b)) and observed an optical response of the circuit on the output 4 using an oscilloscope. This signal allows us to extract the pulse response function of a single phase shifter, which is shown in Fig.~\ref{fig:time}(a). The optical response time $t_{1}$ was measured as a 10-90\% rise time and was found to be different for the heating and the cooling processes: $t_{1}^{h} = 13$~ms and $t_{1}^{c} = 10$~ms.

Next we performed the phase vs. current calibration for each heater to determine the $2\pi$ current thresholds. Fig.~\ref{fig:time}(b) illustrates the intensity dependence on the output port 3 while sweeping the current on the heater 8 in 0.15~mA steps. A heuristic polynomial function $\phi\left(I\right)$ of the form 
\begin{equation}
\phi\left(I\right) = \alpha + \beta I^{2} + \gamma I^{3}
\end{equation}
provides an estimate of the $2\pi$ phase shift current and hence of the dissipated power. The fit of the experimental data provided an estimate of $\beta = (7.45 \pm 0.03)\times 10^{-3}$~rad/mA$^2$. To estimate the thermal crosstalk manifested in parasite phase shifts induced by the neighbouring heaters we measured analogous dependencies while driving the heaters, which ideally should not affect the measured output. For example, Fig.~\ref{fig:time}(c) represents the output intensity pattern in the port~1 with the input light injected in the port~2 while the driving current was swept through the heater~1, which by the design shouldn't modify the output intensity. However, due to a comparable interwaveguide distance $d = 100$~$\mu$m and a heater width of $w = 50$ $\mu$m, the crosstalk effects are almost inevitable in such kind of devices. The crosstalk rate is only several times smaller than $\beta$ on average and should be taken into account.

\begin{figure*}[htbp]
    \centering
    \begin{minipage}{0.3\textwidth}
        \centering
        \includegraphics[width=\linewidth]{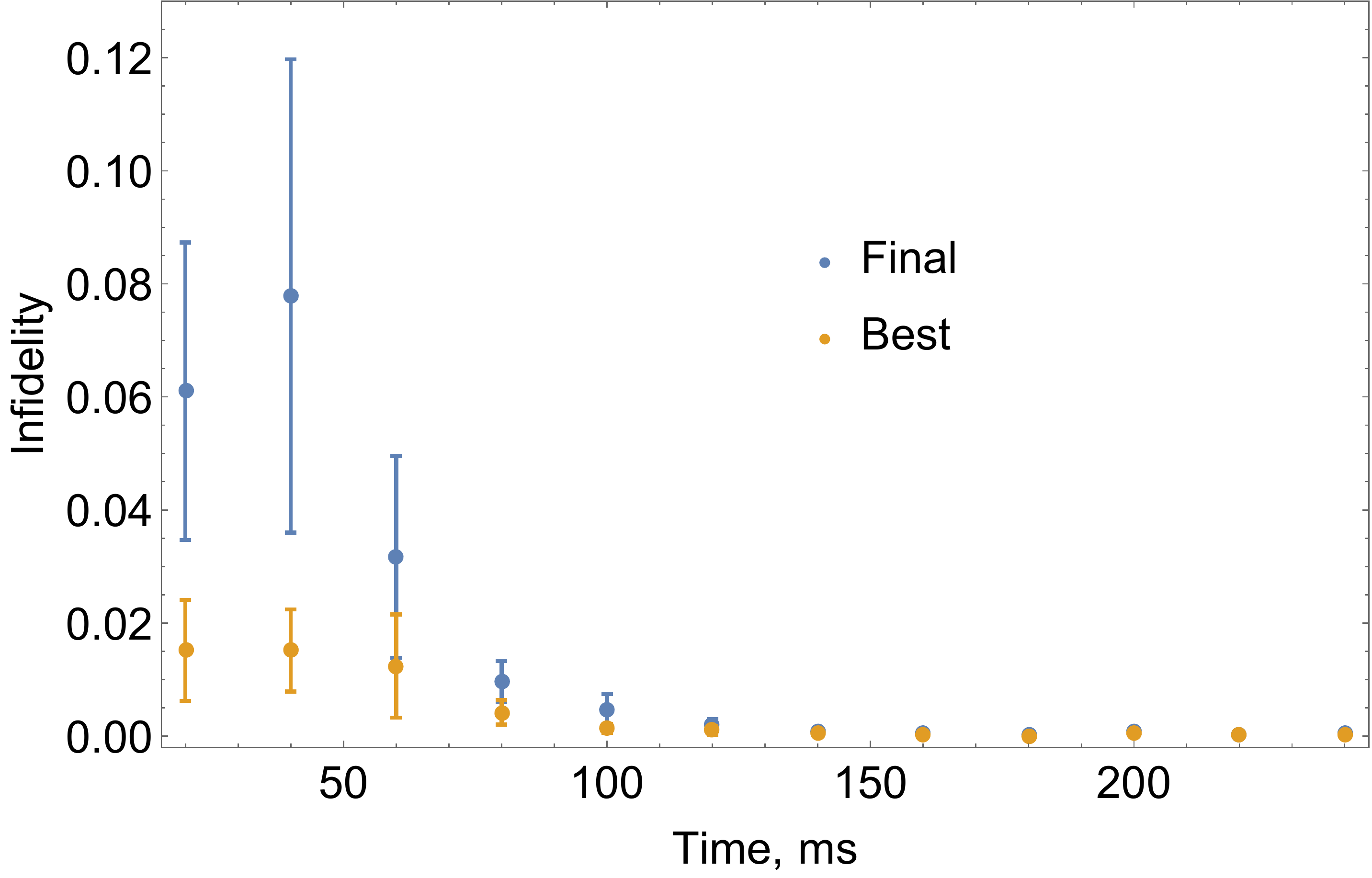}
       (a)
    \end{minipage}\hfill
    \begin{minipage}{0.3\textwidth}
        \centering
        \includegraphics[width=\linewidth]{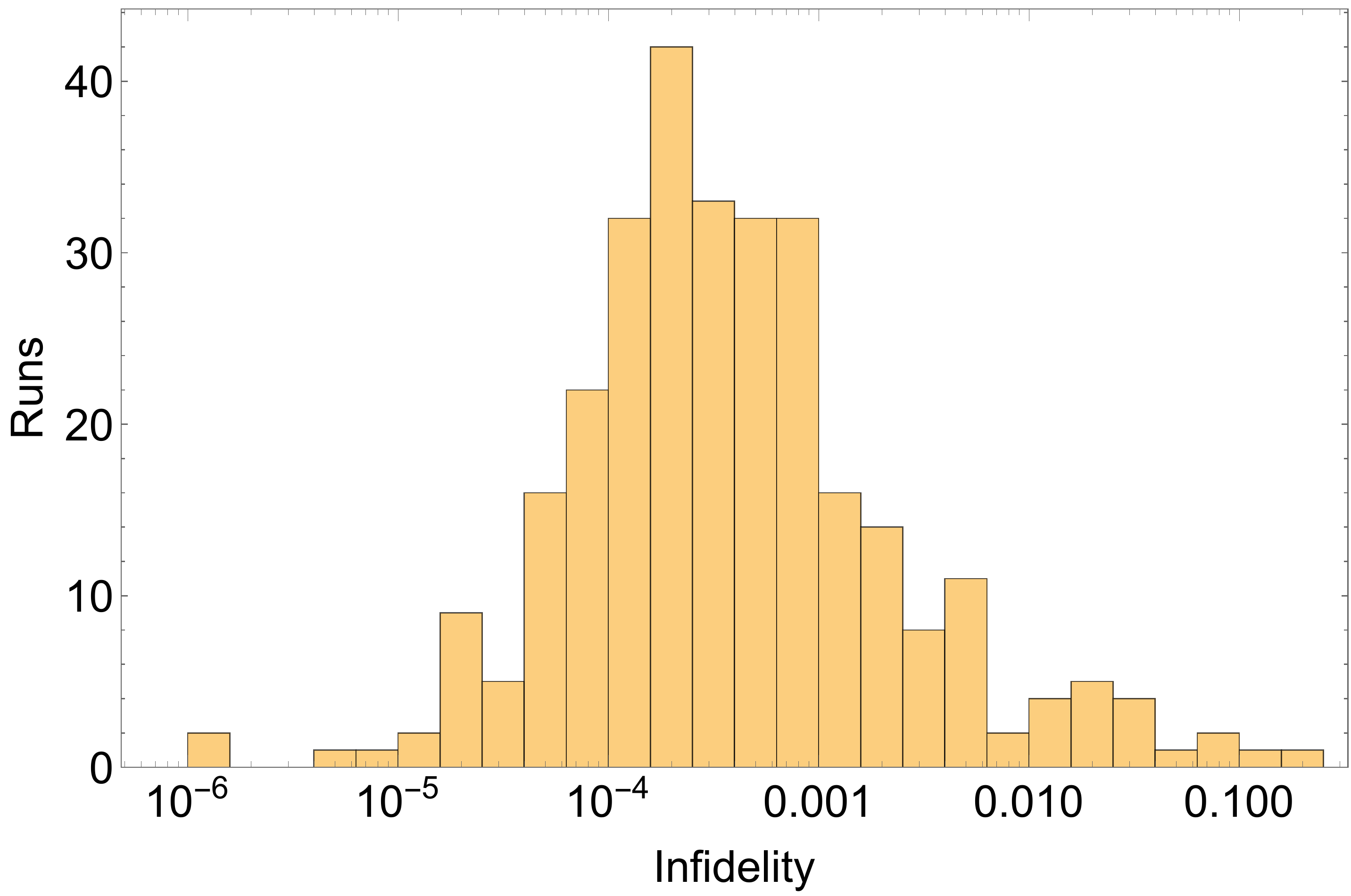}
       (b)
    \end{minipage}\hfill
    \begin{minipage}{0.3\textwidth}
        \centering
        \includegraphics[width=\linewidth]{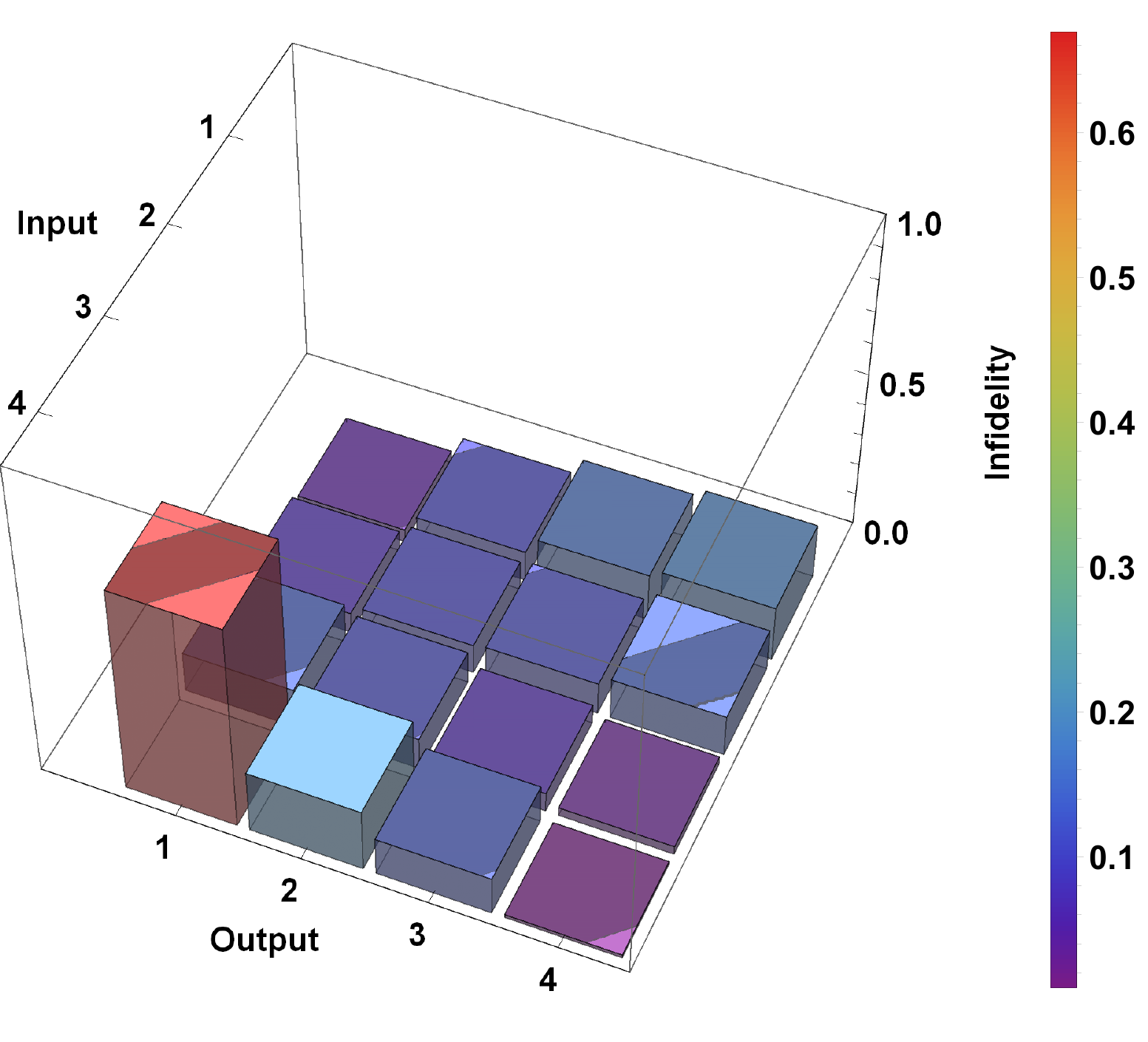}
       (c)
    \end{minipage}
    \caption{(a) The infidelity values for the optimal configuration found during the tuning procedure (blue) and the infidelity values measured after resetting the configuration to the previously found optimal one once the tuning procedure is finished (red). The measurements indicate that even though the single interferometer responds rapidly to the driving current pulse, the whole circuit requires a longer time period to reach the steady state. We used the time delay between the sequential full reconfigurations equal to 180 ms to make sure that each measurement is taken in the thermally stable state. (b) The experimentally measured distribution of the infidelities of the output intensity distributions optimized to the randomly generated target intensity patterns. (c) The infidelity plot of all possible switching configurations of the device. In the worst case scenario (element 1-4) the light is traversing to the most distant output port and hence collecting the effect of more fabrication defects. The infidelity values of all switching configurations are summarized in Tab. \ref{tab::results}.}
    \label{fig:exp_adjustment}
\end{figure*}

\section{Adaptive circuit tuning \label{ACT}}

The multiport interferometer design implemented in our device in principle enables arbitrary unitary transformations in the space of the guided modes \cite{Clements2016}. The integrated photonic chips manufactured with standard lithography process show better performance in terms of precision due to highly repeatable technology of fabricating elementary photonic components, such as directional couplers, which play the central role in the overall precision of universal multiport integrated devices. Hence, the heater calibration provides all the necessary data to program the desired unitary transformation. The FSLW technological process in our laboratory environment provides the repeatability level on a few percent scale which diminishes the performance level and complicates the phase choice decision for the required optical configuration. Thus, in this work we use an adaptive approach to find the pattern of currents through the heating electrodes required to tune the optical circuit to generate the predefined output light distribution. In our experiments we focused on tuning the device solely to reproduce a given intensity distribution between the output ports. Complete characterization of the unitary transformation requires full process tomography and will be reported elsewhere.

The idea of an adaptive thermo-optical circuit adjustment is to launch an optimization procedure, which tunes the phases $\varphi_i$ of the corresponding phase shifters by changing the control currents $I_i$, and step-by-step optimize the output intensity distribution to the desired configuration. We use the following procedure to optimize and test the thermo-optical system performance:
\begin{enumerate}
	\item Laser light is coupled to the $i$-th input port.
	\item Output light transmission $S_j$ is measured for each of the four output ports. Here and further on we use the notation $S_j$ instead of $S_{ij}$ the output intensity, since the input waveguide is not varied during the procedure.
	\item The control voltages are tuned by an optimization algorithm in order to minimize the difference between the measured output intensities $\propto S_j$ and the required ones $\tilde{S_j}$.
	\item The procedure is stopped if the difference function between the desired and the measured intensity configurations is small enough. Otherwise, Steps 2 and 3 are repeated.
\end{enumerate} 

\subsection{Optimization step}
The adjustment procedure consists of two main parts: the calculation of a difference function and an optimization algorithm. These two parts are described below.
\paragraph*{Difference function} The goal of the adjustment procedure is to minimize the difference between the measured and the required normalized output configurations. We used an infidelity function \cite{Nielsen2010} to quantify this difference:
\begin{equation}\label{diff_func}
1-F = 1 - (\sum\limits_{j=1}^{N} \sqrt{S_j \tilde{S_j}})^2.
\end{equation}
The normalized intensities are the quantities, calculated from the experimentally measured output light intensities $I_j$ as:
\begin{equation}
	S_j= \frac{I_j-I_j^{bg}}{\sum\limits_{j=1}^{4}(I_j-I_j^{bg})},
\end{equation}
where $I_j^{bg}$ is the background light intensity in channel $j$.

\paragraph*{Optimization algorithm}
The difference function (\ref{diff_func}) is determined by the 12 control currents as parameters. The global search algorithm should be used for the difference function minimization to avoid convergence to a local minimum. Three search algorithms were tested: simultaneous perturbation stochastic approximation (SPSA) \cite{Spall}, particle swarm optimization (PSO) \cite{Clerc2012} and very fast simulated annealing (VFSA) \cite{Vakil-Baghmisheh2008}. The best search algorithm is supposed to provide the least difference function value given a fixed number of measurements. The VFSA algorithm showed the best performance in numerical simulations (see Appendix \ref{sim}) and was consequently used in the experiment.

First we characterized the time required to reach an equilibrium state $t_2$. For this purpose we ran 300 optimization procedures with a fixed time $t_2$ between the consequent algorithm steps. We set the best found heater currents configuration after each optimization procedure had finished and measured the difference function value. Fig.~\ref{fig:exp_adjustment}(a) illustrates the dependence of the difference function value measured during the optimization search run (best) and after the search run was finished (final). The presented results clearly indicate that even though the switching time of a single Mach-Zender interferometer is fast enough $t_{1} = 10$~ms the whole system of electrodes requires an order of magnitude longer time period to stabilize the temperature distribution across the subsurface volume of the chip. In our setting we used $t_{2} = 180$~ms time delay between the consecutive heater reconfigurations to ensure the steady state output of intensity measurements. This effect is mainly attributed to sufficiently large heater surface areas ($3\times 0.05$~mm heater footprint) required to induce a $2\pi$ optical phase shift.  This is one of the limitations of the FSLW technology, which limits the achievable overlap between the temperature distribution generated by the heater and the optical waveguide core, since writing the waveguides closer than 10~$\mu$m to the top surface of the chip reduces their quality. We intentionally decided to leave the heater sufficiently wide to prevent it from ripping apart while heating due to mechanical stresses induces by laser engraving of the electric pattern on the metal film. However, we believe that the $t_2$ may be sufficiently improved by further adjustments of our fabrication process.

\subsection{Reconfigurability test}

To test the reconfigurability of the device we performed a series of optimization procedures -- tuning the device to operate as a switch, redirecting all the light from a given input port to a given output port, and tuning it to randomly selected target intensity distributions. The results are shown in the Fig.~\ref{fig:exp_adjustment}(c) and Fig.~\ref{fig:exp_adjustment}(b), correspondingly. On average, finding the heater configuration corresponding to a randomly generated pattern turned out to be an easier task in our experimental setting due to the lower sensitivity to the background intensity collected by the output multimode fiber array.\par 
Next we tested the device in the switching regime. 
The switching regime requires setting the intensities in 3 ports as close to zero as possible which is sufficiently limited by the amount of the detected background scattering (mostly due to unguided laser radiation propagating inside the sample). Coupling the chip output to the single-mode fiber array may significantly improve the performance of the optimizer in the switching regime. Fig.~\ref{fig:exp_adjustment}(c) shows the results of optimizing the heater configuration in the intensity switch mode. The worst case scenario for the device under test corresponds to guiding the light towards the most distant output port and results in a sufficient quality reduction. We attribute this behaviour to the impact of the interferometer writing imperfections, i.e. the limited interference visibility for the nested interferometers. For example, the infidelity value $\left(1-F\right)_{41} = 0.669 \pm 0.034$ indicates that the Mach-Zender interferometer controlled with the heater h5 works worse than expected. One of the reasons for such a poor behaviour might be the order of writing of the circuit waveguides and the low feed rate $v = 0.01$~mm/s -- the waveguide with the input 1 and the output 1 was written last in the fabrication sequence and due to the low feed rate the time difference between writing the first and the last waveguide was approximately 8.5~hours. During this period either the exposure conditions could have slightly changed or the dust particles could have contaminated the surface.

\begin{table}[ht!]
	\begin{center}
		\begin{tabular}{ |cc|c|c|c|c| }
			\hline
			&& \multicolumn{4}{c|}{Output} \\
			&& 1 & 2 & 3 & 4 \\ 
			\hline			
			\multirow{4}{*}{\rotatebox{90}{Input}} & 1 & $0.040 \pm 0.006$ & $0.097 \pm 0.017$ & $0.152 \pm 0.018$ & $0.176 \pm 0.003$ \\ 
			& 2 & $0.066 \pm 0.009$ & $0.095 \pm 0.014$ & $0.103 \pm 0.014$ & $0.132 \pm 0.081$ \\ 
			& 3 & $0.129 \pm 0.004$ & $0.103 \pm 0.041$ & $0.062 \pm 0.007$ & $0.027 \pm 0.006$ \\ 
			& 4 & $0.669 \pm 0.034$ & $0.200 \pm 0.041$ & $0.122 \pm 0.042$ & $0.010 \pm 0.002$ \\ 
			\hline
		\end{tabular}
	\end{center}
	\caption{The VFSA algorithm convergence results for the different input/output configurations in the intensity switch mode. The infidelities averaged over 10 runs of the algorithm are shown.}
	\label{tab::results}
\end{table}

\section{Conclusion}

We presented our results on fabrication of a programmable multiport integrated optical circuit with the femtosecond laser writing technology. Our work demonstrates the possibility to produce fully-reconfigurable interferometers with superior switching times -- $10$~ms switching time -- an order of magnitude faster compared to all previously reported results \cite{Flamini2015,Chaboyer2017}. We have characterized the reconfigurablity of the device using the classical input light and developed an adaptive procedure to tune the circuit in accordance with the desired configuration. Further developments should be dedicated to refining the device performance to meet the required precision to operate in quantum regime, i.e. to set arbitrary unitary transformations with high fidelity. This goal requires better localization of the heated chip areas to reduce the crosstalk effects and enhancement of the directional coupler fabrication repeatability. However, even with the current level of performance a reconfigurable circuit of such complexity is a significant step forward for the FSLW technology. We believe that the techniques developed in this work will be used to enable fast and inexpensive fabrication of integrated photonic circuits specifically tuned for particular experiments right in the optical laboratory. Fully reconfigurable integrated circuits are prerequisites for modern experiments in quantum optics and linear-optical quantum computing. Thus the results presented in this work may boost the research in this rapidly developing field.

\begin{acknowledgments}
The work is supported by the Russian Science Foundation project 16-12-00017.
\end{acknowledgments}

\bibliographystyle{apsrev4-1}
\bibliography{bibliography}
\clearpage
\appendix*

\section{Chip geometry}

\begin{figure}
    \centering
    \includegraphics[width=0.45\textwidth]{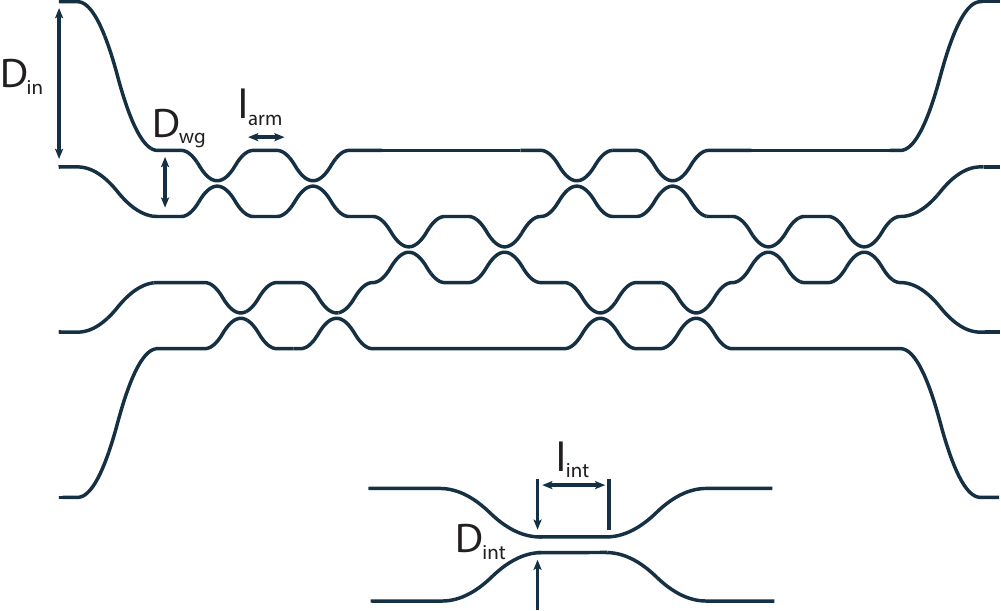}
    \caption{The geometrical parameters of the multiport interferometric integrated circuit fabricated in this work.}
    \label{fig:geometry}
\end{figure}

The geometrical scheme of the integrated multiport interferometer is presented in Fig. \ref{fig:geometry}. The summary of the geometrical parameters of the circuit can be found in table Tab. \ref{tab::geometry}. The curvature radius $R$ denotes the radius of the circular waveguide bends.

\begin{table}[ht!]
\begin{center}
\begin{tabular}{ c|c }
 \hline
 Curvature radius, $R$ & 80 mm \\ 
 Input separation, $D_{in}$ & 250 $\mu$m \\ 
 Waveguide separation, $D_{wg}$ & 100 $\mu$m \\ 
 Interaction separation, $D_{int}$ & 8.57 $\mu$m \\ 
 Interaction length, $l_{int}$ & 0 mm \\
 Arm length, $l_{arm}$ & 2.5 mm \\ 
 \hline
\end{tabular}
\end{center}
\caption{Integrated optical circuit geometry.}
\label{tab::geometry}
\end{table}

\section{Numerical simulations \label{sim}}

Optimization algorithms were tested in numerical simulations. An experimentally obtained value for the difference function may be calculated from the theoretical values of $S_j$:
\begin{equation}
	S_j=S_{ij}=|U_{ij}|,
\end{equation}
where $U$ stands for unitary matrix describing the transformation performed by the chip in the space of optical modes. The unitary $U$ is determined by the thermo-optical circuit phases $\varphi_i$ according to the circuit schematic shown in Fig.~\ref{fig:chip_model}(b). The thermo-optically induced phase delays may be expressed through the controlling currents $I_i$ as follows:
\begin{equation}
	\varphi_i = \sum\limits_{j=1}^{12}\beta_{ij}I_j^2.
\end{equation}
Here we assume that no cross-talk between heaters is present, i.e. $\beta_{ij} \neq 0$ only if $i=j$, and no cubic components $I_i^3$ are used for the simulation purposes. This is justified, since the main goal of the simulation is to tune the search algorithms' parameters for the subsequent usage in the real experiment. That is why the theoretical difference function is just a rough estimate for the experimental one.
  
\begin{figure*}[h!tbp]
	\centering
	\begin{minipage}{0.49\textwidth}
		\centering
		\includegraphics[width=\linewidth]{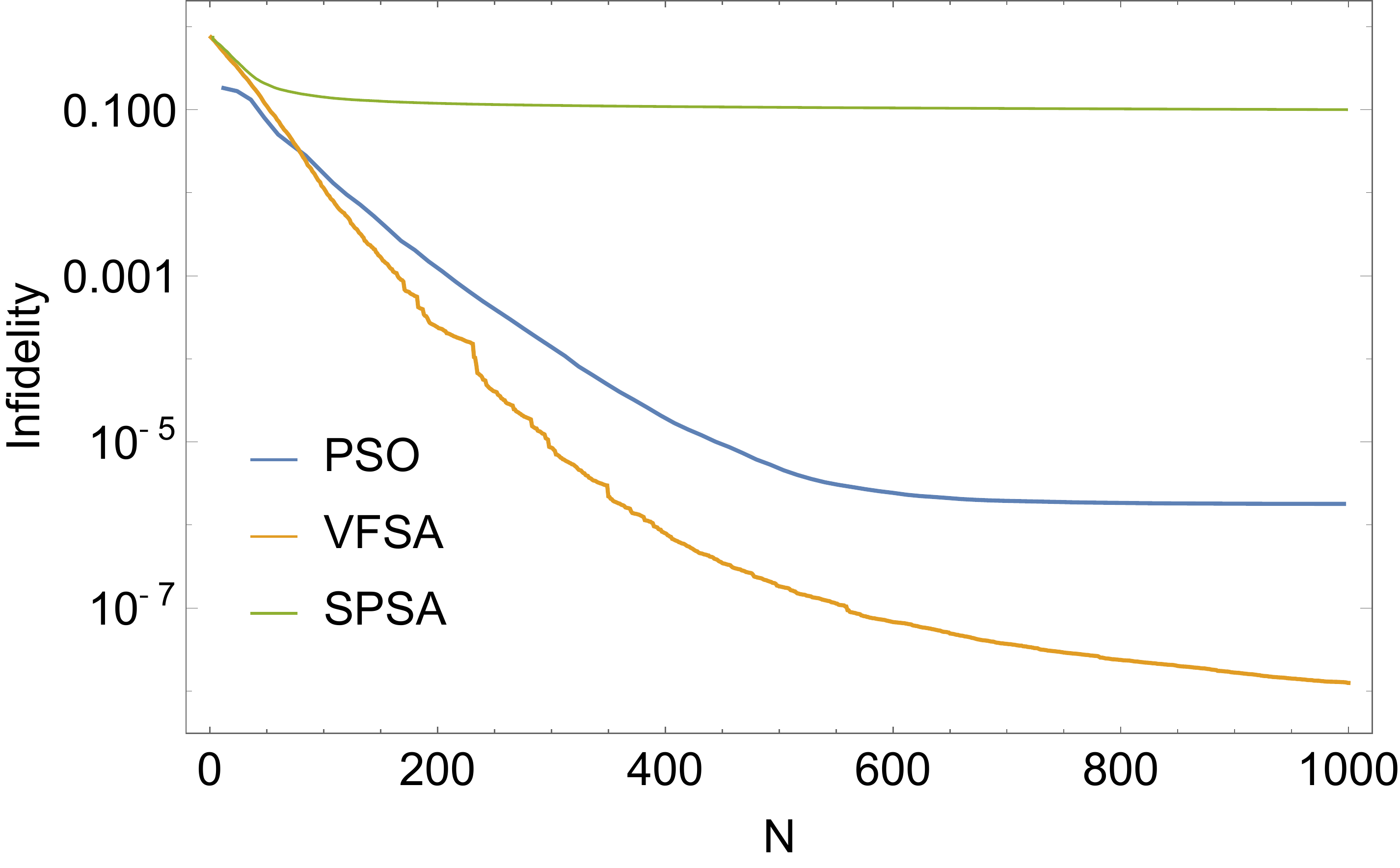}
		(a)
	\end{minipage}\hfill
	\begin{minipage}{0.49\textwidth}
		\centering
		\includegraphics[width=\linewidth]{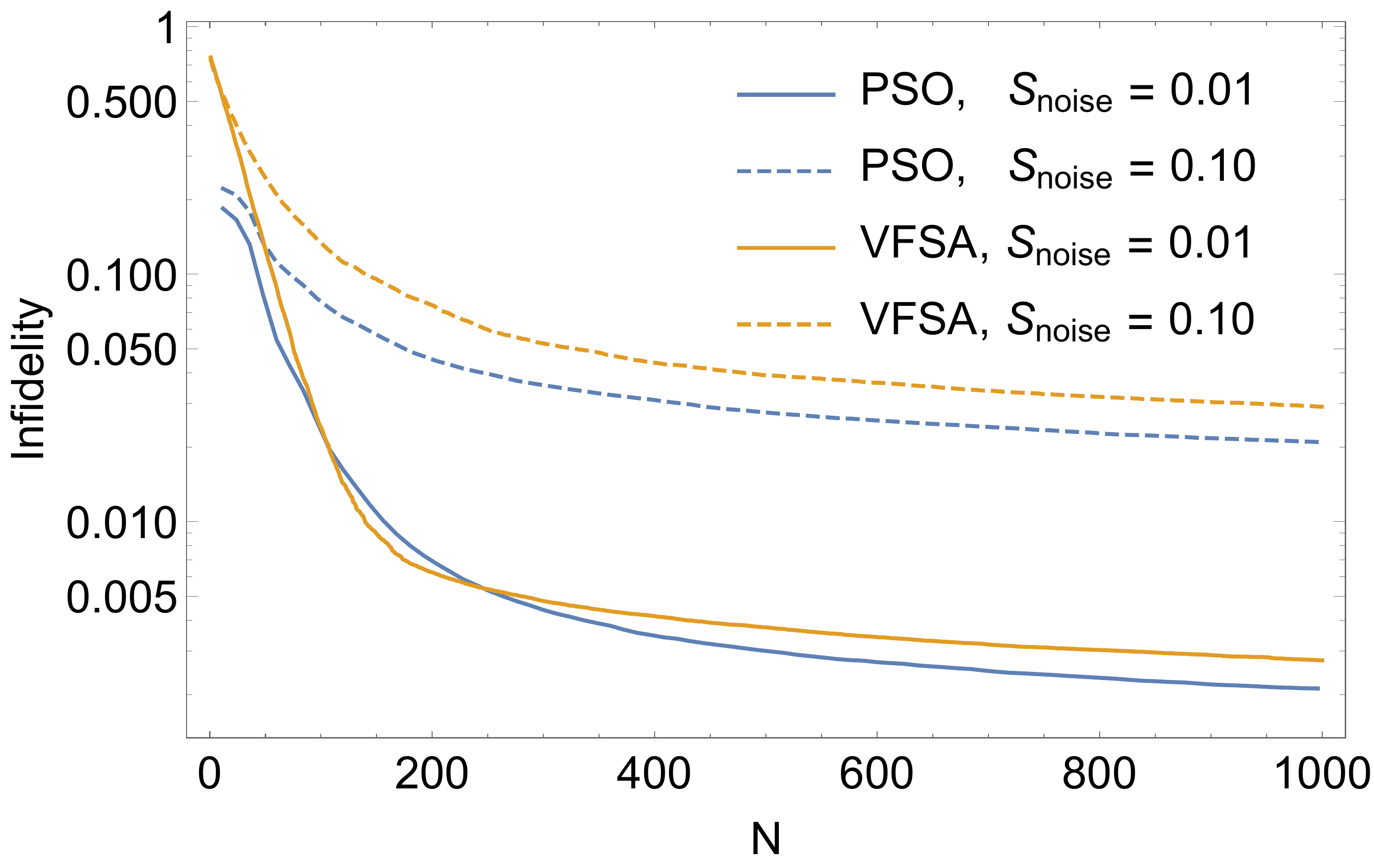}
		(b)
	\end{minipage}
	\caption{Numerical simulations. The convergence of different optimization algorithms to the target configuration $\tilde{S}=\{1,0,0,0\}$. Infidelity is plotted as the function of the number of measurements $N$. Results are averaged over $1000$ runs. (a) No noise   (b) Virtual noise is introduced.}
	\label{fig:simulation}
\end{figure*}

\paragraph*{Noiseless simulations} 
Three optimization algorithms are compared in this section:
\begin{itemize}
    \item Simultaneous perturbation stochastic approximation (SPSA) \cite{Spall}. This algorithm is quite similar to a gradient descent and it is supposed to work well for the possibly noisy functions with few local minima. The SPSA algorithm demonstrated poor performance in comparison to the others, so it was excluded from further consideration. The reasons for the bad performance of SPSA can be probably attributed to sticking to local minima.
    \item Particle swarm optimization (PSO) \cite{Clerc2012}. Although PSO is known to be a global optimization algorithm, the results were generally unsatisfactory. However PSO outperformed other algorithms for "noisy" simulations.
    \item Very fast simulated annealing (VFSA) \cite{Vakil-Baghmisheh2008}. This procedure showed the best performance and was used to acquire the majority of the data in the experiment.
\end{itemize}
Evolution of the theoretical difference function $1-F$ with the number of measurements $N$ for the different optimization algorithms is illustrated in Fig.~\ref{fig:simulation}(a). Results are averaged over $1000$ runs. The target configuration used here was $\tilde{S} = \{1, 0, 0, 0\}$

\paragraph*{Noisy simulations}

Simple experimental noise model was tested to compare the optimization algorithms in non-ideal noisy conditions. It was assumed that the measured value $S_i$ fluctuates in the following way:
\begin{equation}
    S'_i= S_i + \Delta,
\end{equation}
where $\Delta$ is a uniformly distributed random variable $\Delta\in U(-S_{noise},S_{noise})$, here $S_{noise}$ determines the noise value. $S'$ was renormalized subsequently. 

The results of the noisy simulations are given in Fig.~\ref{fig:simulation}(b). PSO  performs better than VFSA under the influence of noise, so both algorithms were tested in the experiment.

\section{Phase calibration details}

\begin{figure*}[h!tbp]
    \centering
    \begin{minipage}{0.49\textwidth}
        \centering
        \includegraphics[width=\linewidth]{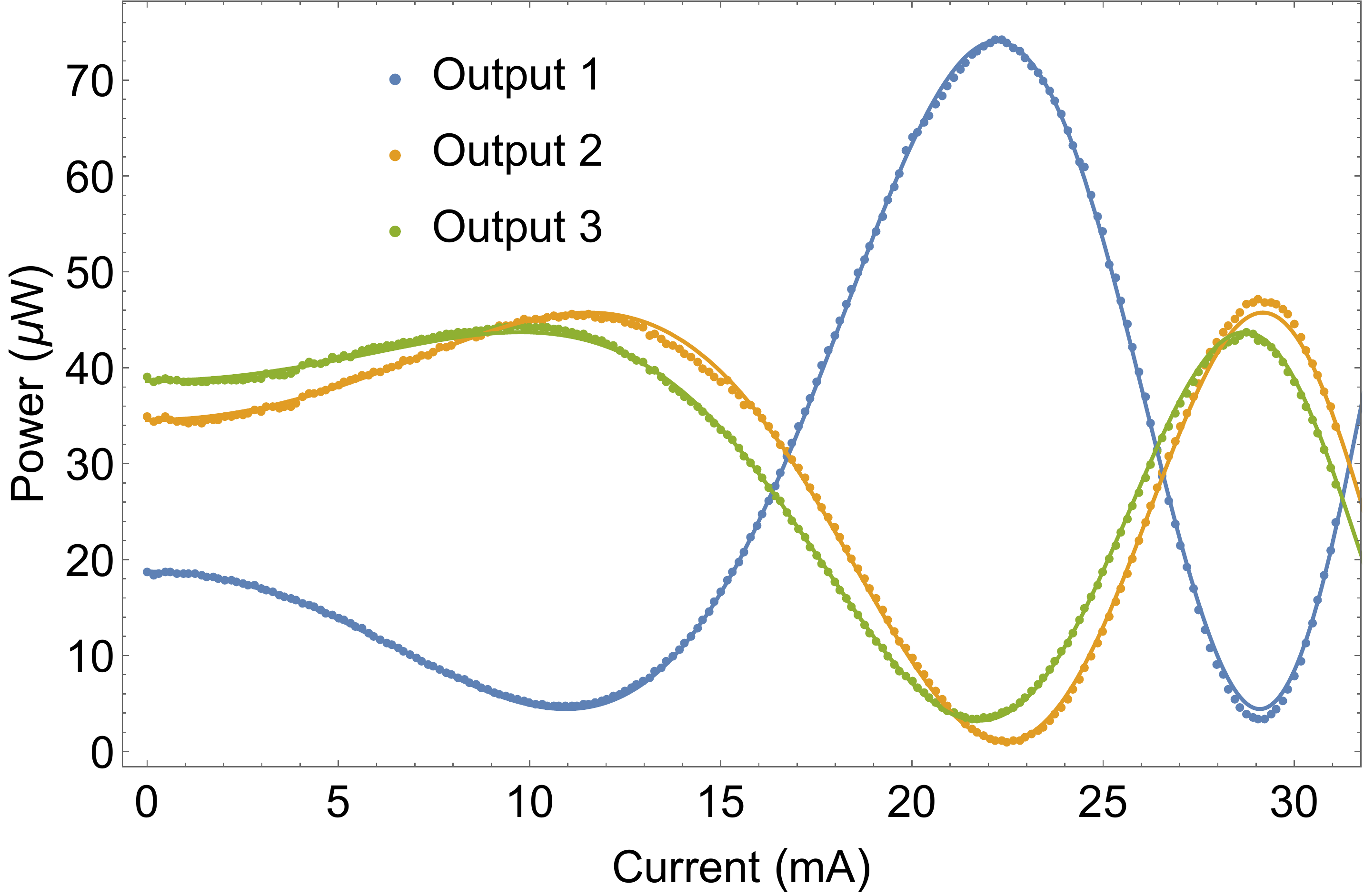}
        (a)
    \end{minipage}\hfill
    \begin{minipage}{0.49\textwidth}
        \centering
        \includegraphics[width=\linewidth]{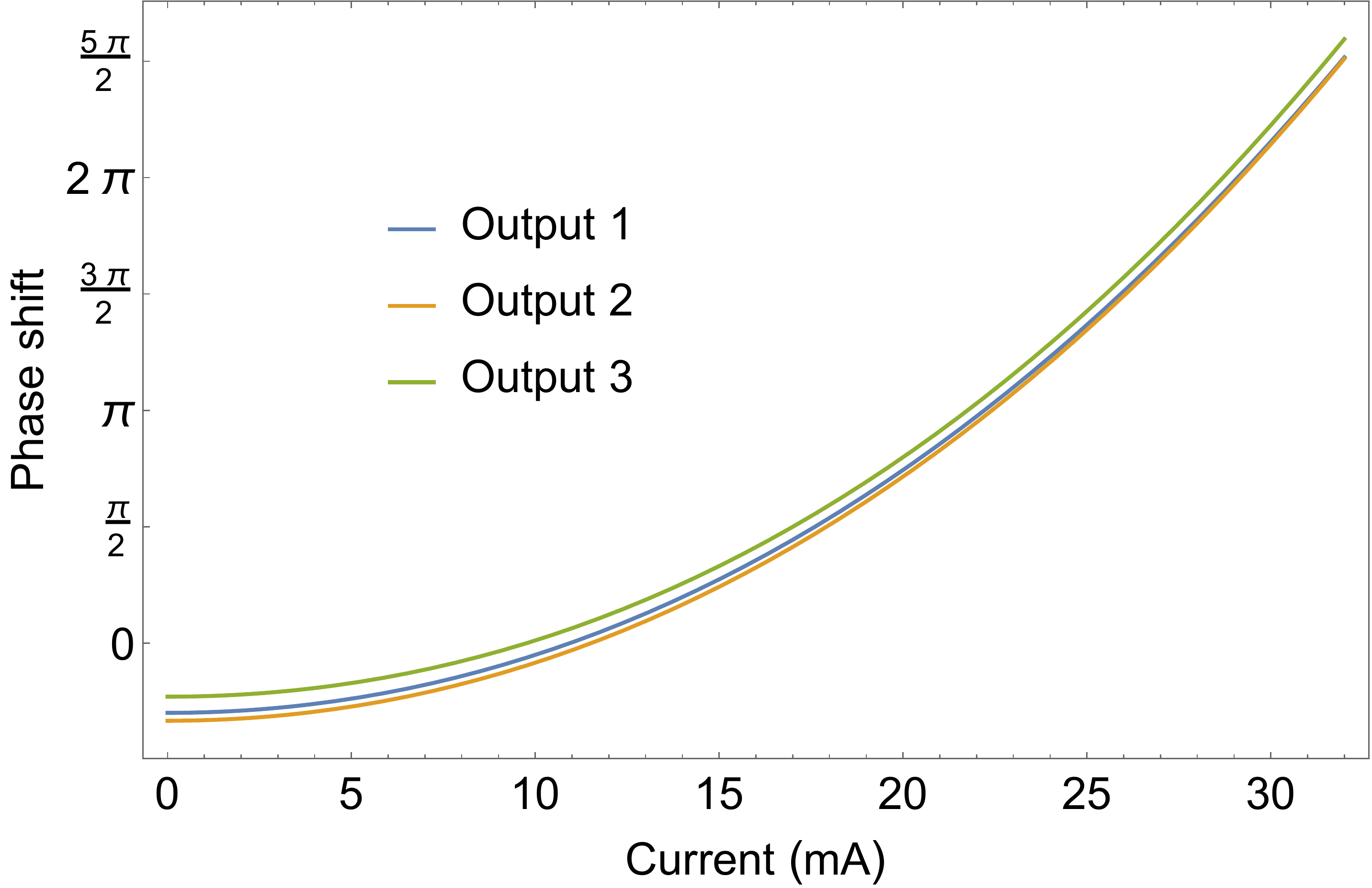}
		(b)
    \end{minipage}
    \caption{ Phase calibration plots. (a) Power in different outputs vs currents through heater h8 were fitted according to model (\ref{model}) to estimate the phase introduced. (b) Phases vs current calibration obtained through different output ports power fitting.}
    \label{fig:calibration_sup}
\end{figure*}

Phases introduced by thermo-optical circuit were fitted according to the model:
\begin{equation}
\phi\left(I\right) = \alpha + \beta I^{2} + \gamma I^{3}.
\label{model}
\end{equation}

Fig. \ref{fig:calibration_sup}(a) represents data used for phase estimation. Different output ports power data for heater h8 were fitted independently, obtained fit parameters are given in Table \ref{tab:exp_calibration}. Final phase vs current calibration curves are in a good agreement (given in Fig. \ref{fig:calibration_sup}(b)).

\begin{table}[ht!]
	\begin{center}
		\begin{tabular}{ |cc|c|c|c| }
			\hline
			&& \multicolumn{3}{|c|}{Parameters} \\
			&& $\alpha$, rad & $\beta, 10^{-3}$ rad/mA$^2$ & $\gamma, 10^{-5}$ rad/mA$^3$ \\ 
			\hline			
			\multirow{3}{*}{\rotatebox{90}{Output}} & 1 & $-0.942 \pm 0.003$ & $7.45 \pm 0.03$ & $3.72 \pm 0.09$  \\ 
			& 2 & $-1.046 \pm 0.005$ & $7.42 \pm 0.06$ & $4.07 \pm 0.19$ \\ 
			& 3 & $-0.724 \pm 0.005$ & $7.10 \pm 0.04$ & $4.90 \pm 0.12$ \\ 
			\hline
		\end{tabular}
	\end{center}
	\caption{Experimental fitting results of heater h8 calibration. Model \ref{model} was used.}
	\label{tab:exp_calibration}
\end{table}

\section{Other experimental optimization results}

\begin{figure*}[h!tbp]
    \centering
    \begin{minipage}{0.49\textwidth}
        \centering
        \includegraphics[width=\linewidth]{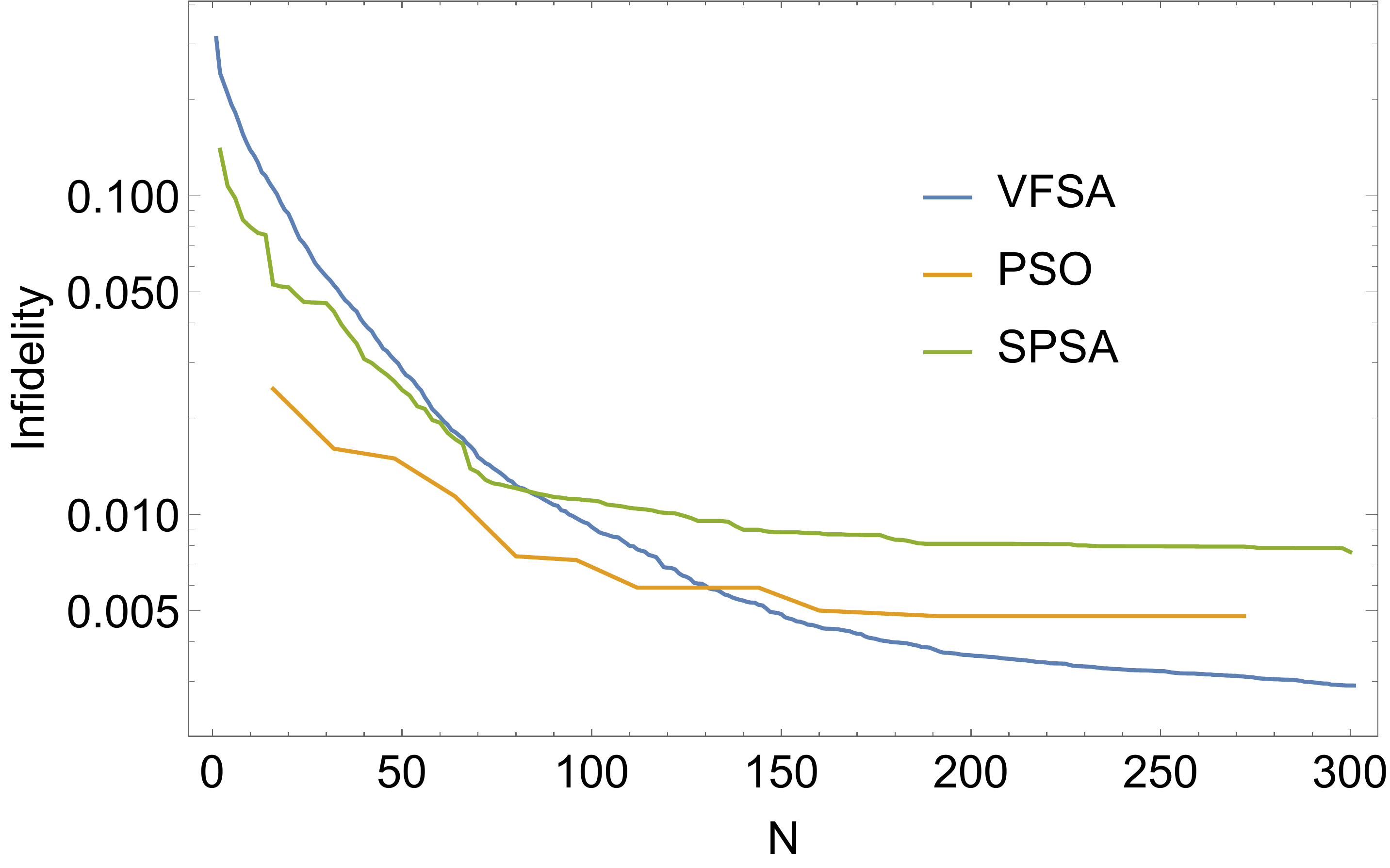}
        (a)
    \end{minipage}\hfill
    \begin{minipage}{0.49\textwidth}
        \centering
        \includegraphics[width=\linewidth]{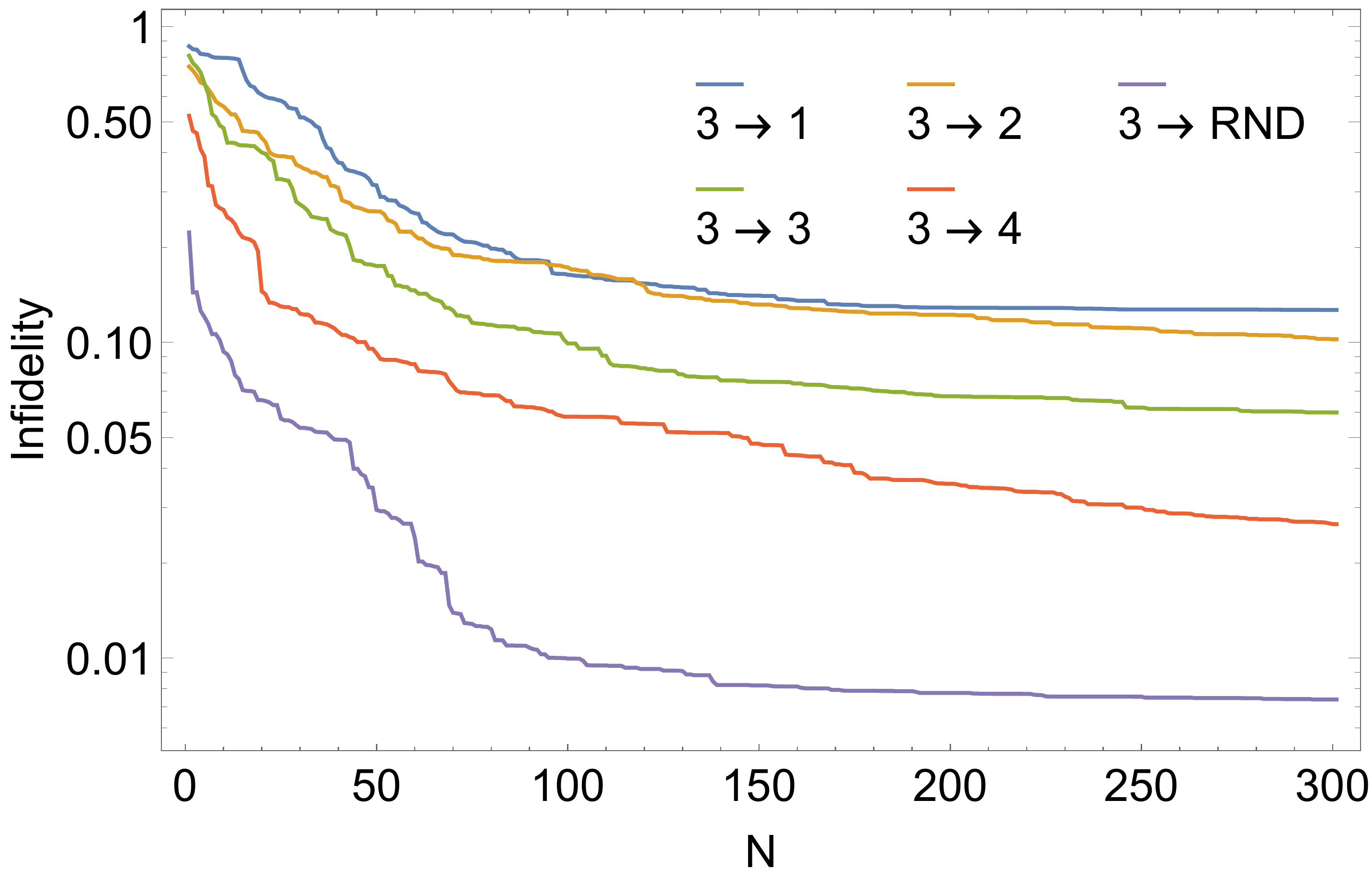}
		(b)
    \end{minipage}
    \caption{ Experimental results averaged over 10 runs. (a) Infidelity dependence on the number of measurements $N$ for the different optimization algorithms. The required output intensity configuration was set to $S_{req}=\{0.25, 0.25, 0.25, 0.25\}$   (b) Convergence to the different output configurations using VFSA. The input port is fixed to port 3. Infidelities are shown for the device operating in switching mode (switching to the output ports $1,\ldots,4$) as well as for convergence to a random output configuration.}
    \label{fig:expetiment_sup}
\end{figure*}

This sections is devoted to some experimental results which are not presented in the section \ref{ACT}.

\paragraph*{Algorithms comparison}

After the preliminary parameter tuning with numerical simulations 3 optimization algorithms were tested in the experiment for the adaptive circuit optimization. The results are shown in Fig.~\ref{fig:expetiment_sup}(a). The VFSA algorithm proved to show the best convergence among the three optimization algorithms described.

\paragraph*{Switching convergence}

Switching the light from the 3rd input port to one of the output posts appeared to be a harder task for the thermo-optical circuit than tuning to a random output configuration $\tilde{S}$. As it is discussed in section \ref{ACT}, the main reason for this is the background light coupled to every output multimode fiber. This light comes from the scattering on the whole structure of the interferometer. This background noise can be reduced by using an SMF fiber array at the output. 

The switching convergence obtained using the VFSA is depicted in Fig.~\ref{fig:expetiment_sup}(b). The background noise spoils the convergence to any configuration which has a $0$ element, e.g. $\tilde{S}=\{0.33, 0, 0.33, 0.34\}$.

\

\end{document}